\documentclass[11pt]{article}
\usepackage{latexsym}  
\usepackage{amssymb}
\usepackage{graphicx}
\usepackage{amsmath}

\begin{document}


\begin{center}
{\Large\bf Can Mass of the Lightest Family Gauge Boson \\
 be of the Order of TeV?}

\vspace{4mm}
{\bf Yoshio Koide}

{\it Department of Physics, Osaka University, 
Toyonaka, Osaka 560-0043, Japan} \\
{\it E-mail address: koide@kuno-g.phys.sci.osaka-u.ac.jp}

\end{center}

\vspace{3mm}
\begin{abstract}
The observed sign of a deviation from the $e$-$\mu$ universality 
in tau decays suggests family gauge bosons with
an inverted mass hierarchy.
Under the constraints from the observed $K^0$-$\bar{K}^0$ and 
$D^0$-$\bar{D}^0$ mixing, we investigate a possibility 
that a mass $M_{33}$ of the lightest gauge boson $A_3^3$ which 
couples with only the third generation quarks and leptons is of 
the order of TeV.  
It is concluded that $M_{33} \sim 1$ TeV is possible if we
adopt a specific model phenomenologically.
\end{abstract}


\vspace{3mm}

\noindent{\large\bf 1. Introduction}

We know three generations of quarks and leptons.
It seems to be natural to regard those as triplets of a family symmetry 
SU(3) \cite{Yanagida78} or U(3). 
However, so far, one has considered that, even if the family gauge 
symmetry exists, it is impossible to observe such gauge boson effects, 
because we know a severe constraint from the observed 
$K^0$-$\bar{K}^0$ mixing \cite{PDG12} and results from $Z'$ search 
\cite{Z_search} at the Tevatron.  
Nevertheless, it is interesting to consider a possibility that 
a family gauge symmetry really exists and the family gauge bosons 
are visible at a lower energy scale.
If there are family gauge bosons, we will inevitably observe the 
deviations from the $e$-$\mu$-$\tau$ universality, although 
whether they are visible or not depends on the breaking scale 
of the family gauge symmetry. 
At present, we know only sizable deviations from the 
$e$-$\mu$-$\tau$ universality in tau decays and upsilon decays
although they are accompanied with large errors, so that they
do not mean violations of the $e$-$\mu$-$\tau$ universality
statistically. 
Nevertheless, they give sufficient curiosity to investigate 
a possibility of family gauge bosons with a lower mass 
scale. 

In this paper, we pay attention to deviations from the 
$e$-$\mu$-$\tau$ universality in the tau decays and in the 
upsilon decays. 
On the other hand, we will give a reconsideration of the 
constraints from the $K^0$-$\bar{K}^0$ and $D^0$-$\bar{D}^0$
mixings. 
Although, we will estimate a mass of the lightest family 
gauge boson from the observed deviations from the $e$-$\mu$-$\tau$ 
universality in this paper, the value is nothing but 
a value for reference, because the experimental values 
have large errors at present.  
(One of the purposes is to call experimental 
physicist's attention to the observation of the deviations 
from the $e$-$\mu$-$\tau$ universality, because they can give 
an important clue to a family gauge boson model, and 
the observations are just within our reach because the data 
have already shown visible deviations.)
We will conclude that a mass of the lightest 
gauge boson $A_3^3$ can be $M_{33} \sim 1$ TeV, if we consider  
a family gauge bosom model with a highly hierarchical mass spectrum.  

The present work has been stimulated by the following observed data 
in the tau decays:
From the present observed branching ratios \cite{PDG12}
$Br(\tau^-\rightarrow \mu^- \bar{\nu}_\mu \nu_\tau) = (17.41 \pm 0.04)\%$ and 
$Br(\tau^-\rightarrow e^- \bar{\nu}_e \nu_\tau) = (17.83 \pm 0.04)\%$, 
we obtain the ratio 
$R_{Br} \equiv Br(\tau^-\rightarrow \mu^- \bar{\nu}_\mu \nu_\tau)/
Br(\tau^-\rightarrow e^- \bar{\nu}_e \nu_\tau) =  0.97644 \pm 0.00314$.
For convenience, we define parameters $\delta_\mu$ and 
$\delta_e$ which are measures of a deviation from 
the $e$-$\mu$ universality as follows:
$$
R_{amp} \equiv \frac{1+\delta_\mu}{1+\delta_e}
= \sqrt{ R_{Br} \frac{f(m_e/m_\tau)}{f(m_\mu/m_\tau)} } 
= 1.0020 \pm 0.0016 ,
\eqno(1)
$$
where $f(x)$ is known as the phase space function and it is given by
$f(x)=1-8 x^2 +8 x^6 -x^8 -12 x^4 \log x^2$.
Then, the result (1) gives
$$
\delta \equiv \delta_\mu -\delta_e = 0.0020 \pm 0.0016 .
\eqno(2)
$$
[The values of the deviation parameters $\delta_\mu$ and $\delta_e$
depend on types of the gauge boson interactions, i.e. 
$(V-A)$, pure $V$, and so on. In Sec.3, we will discuss corrections  
for the parameters $\delta_\mu$ and $\delta_e$ which have been
defined by Eq.(1).]
Of course, from the value (2), 
we cannot conclude that we found a significant difference of 
the deviation from the $e$-$\mu$ universality.  
However, we may speculate a possibility of family gauge bosons.
We can consider that the deviation in the tau decays originates
in exchange of gauge bosons $A_3^2$ and $A_3^1$ which interact
as $\tau \rightarrow A_3^2+ \mu$ and  $\tau \rightarrow A_3^1 + e$, 
respectively, as shown in Fig.1.

\begin{figure}[h]
\begin{picture}(500,70)(-90,0)
  \includegraphics[height=.09\textheight]{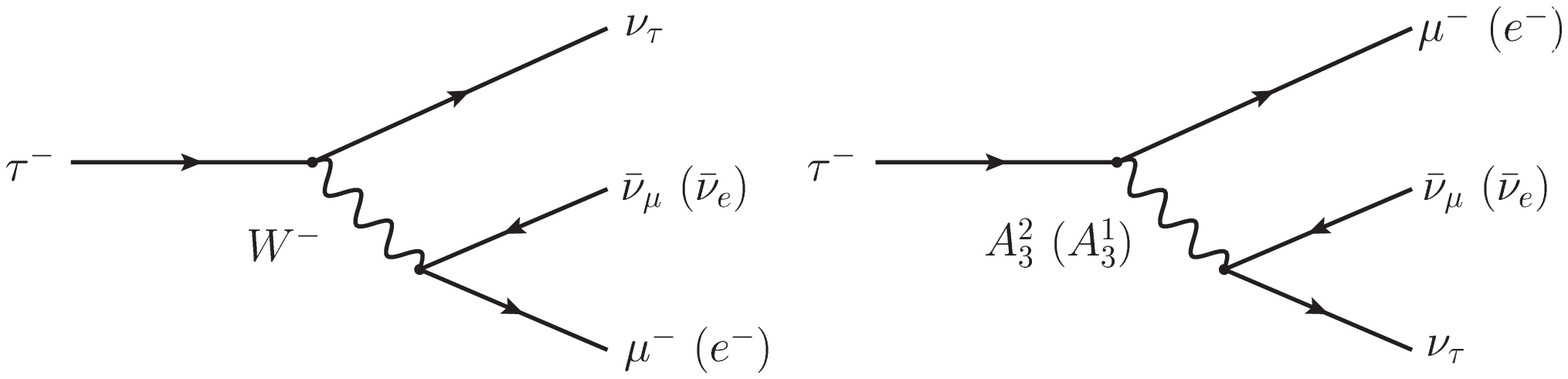}
\end{picture}  
  \caption{Deviation from $e$-$\mu$ universality in tau decays}
  \label{tau_decay}
\end{figure}

Here, let us notice that the observed ratio defined by Eq.(1) shows
$R_{amp} >1$, i.e.  $\delta_\mu >\delta_e$.
Since the deviations are considered as $\delta_i \sim g_F^2/M_{3i}^2$ 
($i=1,2$), this suggests that the mass of $A_3^1$ is larger than that 
of $A_3^2$, i.e.  $M_{31}^2 > M_{32}^2$, where $M_{ij} \equiv m(A_i^j)$.
This suggests that the deviation (1) is caused by family 
gauge bosons with an inverted mass hierarchy. 
(If the gauge boson masses take a normal mass hierarchy, 
we will obtain $\delta_\tau \simeq 0$ because the gauge boson 
$A_3^3$ will take the highest mass.)  
If it is true, the phenomenological aspect for family gauge bosons
will be changed drastically:
(i) A family gauge boson with the highest mass is $A_1^1$, so 
that it is in favor of a relaxation of the severe constraint from
the observed $K^0$-$\bar{K}^0$ mixing. 
(ii) The lightest family 
gauge boson $A_3^3$ interacts with only quarks and leptons of 
the third generation, so that the lightest gauge boson search 
has to be done by $X\rightarrow \tau^+ \tau^-$, 
not by $X\rightarrow e^+ e^-$.  
(The constraint from $Z'\rightarrow \tau^+ \tau^-$ search at 
the Tevatron \cite{Tevatron05} 
cannot be apply to this $A_3^3$ search, because the production rate of
$A_3^3$ is much smaller than that of the conventional $Z'$ boson.) 
(iii) A large deviation from the $\mu$-$\tau$ universality may be
seen in the upsilon decays. 
We consider that it is important to investigate such a possibility 
phenomenologically.

Such a family gauge symmetry model with an inverted mass 
hierarchy has recently been proposed by Yamashita and 
the author \cite{KY_PLB12}. 
In the present paper, we investigate the possibility on the 
basis of this model, 
because mass ratios $M_{ij}/M_{33}$ 
and gauge coupling constant $g_F$ are fixed in the model
(we refer it as Model I)
as we give a brief review in the next section.

We would like to emphasize that, 
in this model, when the family gauge bosons exist
the mass eigenstates, the charged leptons also exist
in the mass eigenstates, while quarks, in
general, do not exist in the mass eigenstates, 
as discussed in Sec.4.
Therefore, the family gauge boson masses are given 
related only to the charged lepton masses, and the
family number violated processes appear only
in the quark sectors, e.g. $\mu \rightarrow e +
\gamma$ is forbidden in the tree-level, while 
$b \rightarrow s + \gamma$ is allowed in the 
tree-level.  
Investigation in the present paper is highly dependent
on the idea in the model \cite{KY_PLB12}.

In Sec.3, we estimate the lightest gauge boson mass $M_{33}$ 
from the tau decay data (1) and also from the upsilon decay data. 
Regrettably, at present, we cannot obtain a conclusive value of
$M_{33}$ because of the large errors. 

Usually, a severe constraint is obtained from the observed 
$K^0$-$\bar{K}^0$ and $D^0$-$\bar{D}^0$ mixing data. 
In Sec.4, we discuss the $K^0$-$\bar{K}^0$ and $D^0$-$\bar{D}^0$
mixings which are caused though 
quark-family mixings $U_d \neq {\bf 1}$ 
and $U_u \neq {\bf 1}$. 
Although the constraint become mild for such a model with
inverse mass hierarchy, it is still severe if 
$(U_d)_{21}$ and $(U_u)_{21}$ are sizable.
Especially, Model I will be ruled out from candidates 
which can interpret both data, i.e. the data in tau and upsilon decays 
and the observed $K^0$-$\bar{K}^0$ and $D^0$-$\bar{D}^0$ mixing data,
because Model I has a mass relation 
$M_{ij} \propto \sqrt{1/m_{ei}+1/m_{ej}}$ and it gives a small ratio 
$M_{22}/M_{33}=4.10$.

In Sec.5, we discuss another models: one has a mass relation 
$M_{ij} \propto (1/m_{ei}+1/m_{ej})$, and the other one has 
a mass relation $M_{ij} \propto 1/m_{ei}m_{ej}$.
The former is a minimum reversion of Model I, but it cannot  
still overcome the constraint from $K^0$-$\bar{K}^0$ mixing
because it gives $M_{22}/M_{33}=16.8$.
The latter can satisfy both constraints because it gives 
$M_{22}/M_{33}=283$, but it is difficult to build a model 
such mass spectrum theoretically.

Sec.6 is devoted to concluding remarks.

\vspace{3mm}

\noindent{\large\bf 2. Family gauge boson model with an inverted mass hierarchy}

The family gauge boson model with an inverted mass hierarchy 
has been proposed stimulated by the Sumino model \cite{Sumino_PLB09}.
Therefore, first, let us give a brief review of the Sumino 
mechanism. 
Sumino has seriously taken why the charged lepton mass formula 
\cite{Koidemass}
$K \equiv ({m_e +m_\mu +m_\tau})/{(\sqrt{m_e} +\sqrt{m_\mu}
+\sqrt{m_\tau})^2} = {2}/{3}$ 
is so remarkably satisfied with the pole masses:
$ K^{pole}=(2/3)\times (0.999989\pm 0.000014)$, 
while if we take the running masses, the ratio becomes
$K(\mu) = (2/3)\times (1.00189 \pm 0.000002)$, for example, 
at $\mu=m_Z$. 
The deviation comes from the $\log m_{ei}^2$ term in the 
QED radiative correction \cite{Arason92}.
Therefore, Sumino has proposed an idea that the factor $\log m_{ei}^2$ 
is canceled by a contribution from family gauge bosons.
In order to work the Sumino mechanism correctly, 
the following conditions are essential:
(i) The left- and right-handed charged leptons $(e_L, e_R)$ 
have to be assigned to $({\bf 3},{\bf 3}^*)$ of the U(3) family 
symmetry, respectively.
(ii) Masses of the gauge bosons are given by 
 $M_{ij}^2  = k( m_{ei} +m_{ej})$.
Thus, the factor $\alpha_{em} \log m_{ei}^2$ due to the photon is 
canceled by a part of 
$-\alpha_F \log M_{ii}^2= -\alpha_F (\frac{1}{2} \log m_{ei}^2 +\log 2k)$ 
due to the family gauge  bosons, where $\alpha_F = g_F^2/4\pi$. 
However, the Sumino model has the following problems:
(i) The model is not anomaly free because the charged leptons are
assigned as $(e_L, e_R)=({\bf 3},{\bf 3}^*)$ of a U(3)$_{fam}$ 
gauge symmetry (this assignment is inevitable in order to the 
so-called Sumino's cancellation mechanism \cite{Sumino_PLB09}); 
(ii) Effective current-current interactions with $\Delta N_{fam} =2$ appear
because of the $(e_L, e_R)=({\bf 3},{\bf 3}^*)$ assignment; 
(iii) The Sumino's cancellation mechanism cannot be applied 
to a SUSY model,
because the vertex type diagram does not work in a SUSY model. 

Therefore, in order to evade the above problems, 
in the revised model \cite{KY_PLB12}, we assign the U(3)$_{fam}$ 
quantum numbers as $(e_L, e_R)=({\bf 3},{\bf 3})$, so that
the model is anomaly free, and 
the $\Delta N_{fam} =2$ interactions do not appear at tree level.
On the other hand, in order to realize the cancellation 
mechanism, we must consider that masses $M_{ij}$ of 
the gauge bosons $A_i^j$ are given as follows:
$$
m^2(A_i^j) \equiv M_{ij}^2 = k \left( \frac{1}{m_{ei}} +
\frac{1}{m_{ej}} \right) ,
\eqno(3)
$$
differently from those in the Sumino model, 
$M_{ij}^2 = k (m_{ei}+m_{ej})$, 
where $m_{ei}$ are charged lepton masses.
(Note that $\log M_{ii}^2 = +\frac{1}{2}\log m_{ei}^2 + \log 2k$ in the 
Sumino model, while $\log M_{ii}^2 = -\frac{1}{2}\log m_{ei}^2 + \log 2k$
in our model). 

As well as the Sumino model, the family gauge coupling constant $g_F$
in our model is not a free parameter because the cancellation 
mechanism: 
$$
g_F^2 = \frac{3}{2} \zeta\ e^2 = \frac{3}{2} \zeta\ g_W^2 \sin^2 \theta_W ,
\eqno(4)
$$
where $g_W$ is the weak gauge coupling constant given by 
$G_F/\sqrt{2} = g_W^2/8 M_W^2$, and $\zeta$ is a fine tuning parameter.
In our model, the parameter $\zeta$ 
is numerically given by $\zeta=1.752$ ($\zeta \simeq 7/4$)
\cite{KY_PLB12}.
(Hereafter, in numerical estimates of $g_F$, we will use input values 
$\zeta=7/4$ and $\sin^2 \theta_W=0.223$.) 
Only a free parameter in the model is the magnitude of $M_{33}$
because the ratios $M_{ij}/M_{33}$ are fixed by the relation (3):
$M_{33} : M_{23} :  M_{22} : M_{13} : M_{12} : M_{11} =
1 : 2.98 : 4.10 : 41.70 : 41.80 : 58.97$. 

The family gauge boson interactions are given by
$$
H_{fam} = g_F (\bar{e}^i \gamma_\mu e_j) (A^\mu)_j^i ,
\eqno(5)
$$
because the U(3) triplet assignment for charged leptons 
is given by $(e_L, e_R)=({\bf 3},{\bf 3})$ which gives
anomaly free configuration. 
Note that the interaction type is pure vector differently from that in 
the Sumino model, in which the currents have been given by $(V \pm A)$. 
(For example, a decay $B_s^0 \rightarrow \tau^- +\mu^+$ via 
an exchange of family gauge boson $A_3^2$ is forbidden. )

Note that the family gauge bosons are in the mass-eigenstates 
on the flavor basis in which the charged lepton mass matrix is
diagonal.
In this model, a lepton number violating process never occurs 
at the tree level of the current-current interaction in the
charged leptons. 
As we discuss in Sec.4, since quarks are not in the mass-eigenstates
on the diagonal basis of the charged lepton mass matrix, 
family number changing interactions appear in the 
quark-quark and quark-lepton interactions.
For example, the $K^0$-$\bar{K}^0$ mixing is cased only 
through the quark mixings. 
The $\mu$-$e$ conversion $\mu^- +N \rightarrow e^- +N$ is
also caused through the quark mixings.


\vspace{3mm}

\noindent{\large\bf 3.  Mass of the lightest gauge boson}

First, on the basis of the model with the gauge boson masses (3), 
we investigate 
a possible deviation from the $e$-$\mu$ universality in 
the tau decays, because the processes are pure leptonic, so that
they are not affected by quark family mixing. 
(Although the estimate was already discussed in Ref.\cite{KY_PLB12}, 
the purpose was only to estimate an order of the energy scale
roughly, and the relation (4) was not used.) 
In the present model, the deviation from 
the $e$-$\mu$ universality is characterized by the parameters
$$
\delta^0_i= \frac{g_F^2/M_{3i}^2}{g_W^2/8M_W^2} ,
\eqno(6)
$$
where $i=1,2$ (i.e. $i=e, \mu$) in the tau decays. 
Since $(M_{32}/M_{31})^2 = 0.00508$ from the relation (3),
we neglect the contribution $\delta_e^0$ compared with
the contribution $\delta_\mu^0$ hereafter. 
Since the interactions (5) with 
the family gauge bosons are pure vector, our 
parameter $\delta_\mu^0$ does not directly mean
the observed $\delta_\mu$. 
The effective four Fermi interaction for $\tau^- \rightarrow 
\mu^- \bar{\nu}_\mu \nu_\tau$ is given by
$$
H^{eff} = \frac{G_F}{\sqrt{2}} \left\{ 
[\bar{\mu} \gamma_\rho (1-\gamma_5) \nu_\mu]
[\bar{\nu}_\tau \gamma^\rho (1-\gamma_5) \tau]
+ \delta_\mu^0 (\bar{\nu}_{L\tau} \gamma_\rho \nu_{L\mu})
(\bar{\mu} \gamma^\rho \tau ) \right\} ,
\eqno(7)
$$
where we have dropped the term 
$(\bar{\nu}_{R\tau} \gamma_\rho \nu_{R\mu})$ 
because $\nu_R$ have large Majorana masses. 
By using Fierz transformation, we can express Eq.(7) 
as
$$
H^{eff} = 4 \frac{G_F}{\sqrt{2}} \left\{ 
\left( 1+ \frac{1}{4} \delta_\mu^0 \right) 
(\bar{\mu}_L \gamma_\rho \nu_{L\mu})
(\bar{\nu}_{L\tau} \gamma^\rho \tau_L) 
- \frac{1}{2} \delta_\mu^0 
(\bar{\mu}_R \nu_{L\mu}) (\bar{\nu}_{L\tau} \tau_R) 
 \right\} .
\eqno(8)
$$
Therefore, the observed $\delta_\mu$ is 
related to our parameter $\delta_\mu^0$ as follows:
$$
\delta_\mu = \frac{1}{2} \left( 1 - 2 x_\mu \frac{g(x_\mu)}{f(x_\mu)}
\right) \delta_\mu^0 ,
\eqno(9)
$$
where $g(x)=1+ 9 x^2 - 9 x^4 -x^6 +6 x^2(1+x^2) \log x^2$ and  
$x_\mu = m_\mu/m_\tau$. 
Here, we have neglected higher terms of $\delta_\mu^0$. 
(For more details, for example, see Ref.\cite{Pich-Silva_PRD95}.)
The present deviation $\delta \equiv \delta_\mu -\delta_e
= (2.0 \pm 1.6)\times 10^{-2}$ gives a family gauge boson mass of $A^2_3$
$$
M_{23} =2.6^{+3.2}_{-0.7} \ {\rm TeV}.
\eqno(10)
$$
so that it means  the lightest family 
gauge boson mass 
$$
M_{33} =0.87^{+1.07}_{-0.22} \ {\rm TeV},
\eqno(11)
$$
from the mass relation (3). 
However, at present, the numerical result (10) [also (11)] should not be taken 
rigidly, because, for example, if we change the input value $\delta$ 
from the input value $\delta=0.0020 \pm 0.0016$ to 
$\delta = 0.0020 \pm 0.0016\times 1.25$, the predicted upper value
of $M_{33}$ will become $M_{33} \rightarrow \infty$.

\begin{figure}[h]
\begin{picture}(500,50)(-30,-10)
  \includegraphics[height=.08\textheight]{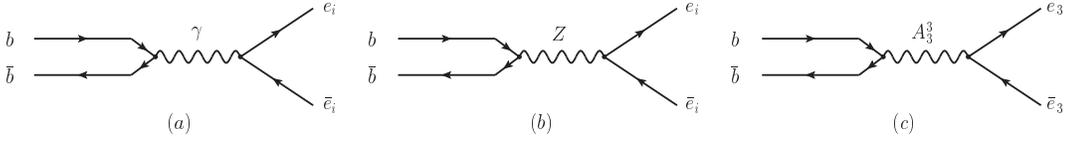}
\end{picture}  
  \caption{Deviation from $e$-$\mu$-$\tau$ universality in upsilon decay}
  \label{upsilon decay}
\end{figure}

At present we have another data of deviations from the 
$e$-$\mu$-$\tau$ universality, i.e. 
data of upsilon decays 
$\Upsilon(1S) \rightarrow \ell^+ \ell^-$  ($\ell = e, \mu, \tau$).
For the time being, we neglect family mixing among quark families.
Then, the $b\bar{b}$ sector couples only to the family gauge boson 
$A_3^3$ in addition to the standard gauge bosons (photon and 
$Z$ boson) as seen in Fig.2. 
Present experimental data \cite{PDG12} 
$Br(\Upsilon(1S)\rightarrow \tau^+\tau^-) = (2.60 \pm 0.10) \%$,   
$Br(\Upsilon(1S)\rightarrow \mu^+\mu^-) =(2.48 \pm 0.05 )\%$, and
$Br(\Upsilon(1S)\rightarrow e^+ e^-) = (2.38 \pm 0.11) \%$ 
gives $R_{Br} \equiv Br(\Upsilon \rightarrow \tau^+\tau^-)/
Br(\Upsilon \rightarrow \mu^+\mu^-)  = 1.048 \pm 0.046$,
which leads to
$$
R_{amp} = 1+\delta_{\tau/\mu} = 1.028 \pm 0.022 , 
\eqno(12)
$$
where $R_{amp}$ has been defined by
$R_{amp} \equiv  \sqrt{ R_{Br}/ R_{kine} }$,
$$
R_{kine}^{\tau/\mu} = \frac{1+2 \frac{m_\tau^2}{M^2} }{
1+2 \frac{m_\mu^2}{M^2}}
\sqrt{ \frac{1-4\frac{m_\tau^2}{M^2} }{1-4\frac{m_\mu^2}{M^2} } }.
\eqno(13)
$$
Also, we obtain $R_{amp}^{\mu/e}= 1+ \delta_{\mu/e} =1.021 \pm 0.051$. 
However, hereafter, we will not utilize the data on 
$Br(\Upsilon(1S)\rightarrow e^+ e^-)$ because of its large error. 
Since the contributions from photon, $Z$ boson, and $A_3^3$ boson,
are characterized by $1/q^2$, $1/(q^2-M_Z^2)$ and $1/(q^2-M_{33}^2)$
with $q^2 = M_\Upsilon^2$,
respectively, the sign of the deviation $\delta_\tau$ has to
be negative considering naively, while the observed result (12) 
has denoted that it is positive. 
Therefore, we assume that quark fields are assigned as 
$(q_L, q_R) \sim ({\bf 3}^*,{\bf 3}^*)$ of the U(3) family symmetry,
differently from that in the charged lepton sector, 
$(e_L, e_R) \sim ({\bf 3},{\bf 3})$. 
[The model is still anomaly free in spite of this modification, 
differently from the Sumino model with $({\bf 3}, {\bf 3}^*)$.]
Since we can neglect the $Z$ boson contribution compared with 
the photon contribution,
the deviation parameter $\delta_\tau$ is given
$$
\delta_\tau = \frac{g_F^2}{e^2/3} \frac{M_\Upsilon^2}{M_{33}^2},
\eqno(14)
$$
where the factor $1/3$ has originated in the electric charge of 
$b$ quark.
The observed deviation $\delta_\tau =0.028 \pm 0.022$ gives
$$
M_{33} =(112^{+130}_{\ -26}) \  {\rm GeV} . 
\eqno(15)
$$
This value is considerably small compared with the result (11) from
the tau decay data. 
However, the upper bound of $M_{33}$ is sensitive to the input 
value of $\delta_\tau$.

Although we cannot obtain a conclusive value of $M_{33}$ after all,
it should be noted that those results show that the determination 
of $M_{33}$ is within our reach. 
We hope to obtain more accurate data on the deviation from 
$e$-$\mu$-$\tau$ universality in near future.

\vspace{3mm}

\noindent{\large\bf 4.  Family-number violating processes due to quark mixing}

So far, we have not discussed family mixing in the quark sectors.
In the present model, the family number is defined by a flavor basis 
in which the charged lepton mass matrix $M_e$ is diagonal, while, 
in general, quark mass matrices $M_u$ and $M_d$ are not diagonal 
in this basis.
When we denote quarks in the mass eigenstates as $u=(u,c,t)$
and $d=(d, s, b)$,  and those in the family eigenstates as 
$u^0=(u_1^0,u_2^0,u_3^0)$ and $d^0=(d_1^0,d_2^0,d_3^0)$, 
the family mixing matrices are defined as $q^0_{L} = U^q_{L} q_{L}$
(and also $L\rightarrow R$)  ($q=u, d$). 
Quark mass matrices $M_q$ are diagonalized as 
$(U^q_L)^\dagger M_q U^q_R = D_q$, and the quark mixing matrix
$V_{CKM}$ \cite{CKM} is given by $V_{CKM} = (U^u_L)^\dagger U_L^d$. 
[Since $(\nu_i, e^-_i)_L$ are doublets in SU(2)$_L$, 
we can regard the eigenstates of the family symmetry as 
the eigenstates of weak interactions.] 
Since we know $V_{CKM} \neq {\bf 1}$, we cannot take 
the mixing matrices $U^u_L$ and $U^d_L$ as $U^u_L={\bf 1}$ and 
$U^d_L={\bf 1}$ simultaneously.  
Under this definition of the mixing matrices, the family 
gauge bosons interact with quarks as follows:
$$
H_{fam} = g_F \sum_{q=u,d} (\bar{q^0}^i \gamma_\mu q^0_j) (A^\mu)_i^j
=  g_F \sum_{q=u,d} (A_\mu)_i^j  \left[ (U_L^{q *})_{i k}
(U_L^q)_{j l} (\bar{q}_{L k} \gamma^\mu  q_{L l}) +
(L \rightarrow R) \right] .
\eqno(16)
$$ 

In the investigation of the upsilon decays in Sec.3, 
we may consider that $b$-$s$ mixing (i.e. $U^d_{31}$ and $U^d_{32}$) is 
highly suppressed, considering the observed CKM mixing 
$|V_{ub}| \sim 10^{-3}$ and $|V_{cb}| \sim 10^{-2}$. 
If the mixing is sizable, we would observe a decay 
$\Upsilon \rightarrow \mu^\pm \tau^\mp$ 
(the data \cite{PDG12} show $Br( \Upsilon \rightarrow \mu^\pm \tau^\mp)
< 6.0 \times 10^{-6}$). 
We can consider that the estimate in Eq.(12) with neglecting the 
$b$-$s$-$d$ mixing is reasonable.

The greatest interest to us is whether we can take a lower  
value of $M_{33}$ without contradicting the constraint
from the observed $K^0$-$\bar{K}^0$ mixing. 
The $K^0$-$\bar{K}^0$ mixing is caused by $A_1^1$, $A_2^2$
and $A_3^3$ exchanges only when the down-quark mixing 
$U^d_{L/R} \neq  {\bf 1}$ exists:
$$
H^{eff} = g_F^2 \left[ 
\frac{1}{M_{33}^2} (U^{d *}_{31} U^d_{32})^2 +
\frac{1}{M_{22}^2} (U^{d *}_{21} U^d_{22})^2 +
\frac{1}{M_{11}^2} (U^{d *}_{11} U^d_{12})^2 \right] 
(\bar{s}\gamma_\mu d) (\bar{s}\gamma^\mu d) + h.c.,
\eqno(17)
$$
where, for simplicity, we have taken $U_L^d =U_R^d$.  
If we assume the vacuum-insertion approximation,
we obtain 
$$
\Delta m_K^{fam} = \left[ \frac{1}{M_{33}^2} (U^{d *}_{31} U^d_{32})^2 +
\frac{1}{M_{22}^2}(U^{d *}_{21} U^d_{22})^2  +
\frac{1}{M_{11}^2}(U^{d *}_{11} U^d_{12})^2 \right] 
\times {0.7738 \times 10^{-12} } 
\ {\rm [TeV]} ,
\eqno(18)
$$
where the value of $M_{33}$ is taken in a unit of TeV, 
and we have used values $f_K =0.1561$ GeV, 
$m_s(0.5{\rm GeV})=0.513$ GeV and $m_d(0.5{\rm GeV})=0.0259$
GeV.
On the other hand, the observed value \cite{PDG12}
is  $\Delta m_K = (4.484 \pm 0.006) \times 10^{-18}$
TeV, and the standard model has a share of 
$\Delta m_K \sim 2 \times 10^{-18}$ TeV (for example, 
see Ref.\cite{Bigi-Sanda}, and for recent work, for instance, see
the second one in Refs.\cite{Yanagida78}). 

We know the observed values of $V_{CKM}$ parameters 
\cite{PDG12}, but we do not know the mixing values 
$U^d$ and $U^u$ separately.
By way of trial, let us take 
$U^d=U(\frac{1}{2}\theta_{12}, \frac{1}{2}\theta_{23}, 
\frac{1}{2}\theta_{13}, \delta_d)$ 
and $U^u =U(-\frac{1}{2}\theta_{12}, -\frac{1}{2}\theta_{23}, 
-\frac{1}{2}\theta_{13}, \delta_u)$ corresponding to the standard 
expression of the CKM matrix $V_{CKM}=U(\theta_{12}, \theta_{23},
\theta_{13}, \delta)$ with $\theta_{12}=13.02^\circ$, 
$\theta_{23}=2.36^\circ$, $\theta_{13}=0.201^\circ$ and 
$\delta=69.0^\circ$.
In order to reproduce the observed $V_{CKM}$, we must take 
$\delta_u \sim 80^\circ$ and $\delta_d \simeq 25^\circ$.  
This tentative choice gives
$$
|U^d| = \left(
\begin{array}{ccc}
0.9935 & 0.1134 & 0.00176 \\
0.1133 & 0.9933 & 0.02060 \\
0.003985 & 0.02029 & 0.9998 
\end{array} \right) , \ \ \ 
|U^u| = \left(
\begin{array}{ccc}
0.9935 & 0.1134 & 0.00176 \\
0.1134 & 0.9933 & 0.02060 \\
0.00266 & 0.02051 & 0.9998 
\end{array} \right) . \ \ \ 
\eqno(19)
$$
Of course, there is a possibility that $U^u$ and $U^d$
take large mixings with opposite signs each other,
e.g. $|\theta^d_{12}| \gg |\theta_C|$ and 
$|\theta^u_{12}| \gg |\theta_C|$, but 
$\theta^d_{12} -\theta^u_{12} = \theta_C$
($\theta_C$ is the Cabibbo angle).
Such a case will give not only large mixings
in $K^0$-$\bar{K}^0$, $D^0$-$\bar{D}^0$, $\cdots$ systems, 
but also large rates of $\mu$-$e$ conversion 
($\mu N \rightarrow e N$) and $s \rightarrow d +\gamma$.
In this paper, we do not consider such a case,
and we consider a case $U^d \sim U^u \sim V_{CKM}$
with small mixings.

If we take the mixing $U^d$ given in Eq.(19) on trial, 
from $|U^{d *}_{31} U^d_{32}|^2=6.539 \times 10^{-9}$,
$|U^{d *}_{21} U^d_{22}|^2=0.1268$ and 
$|U^{d *}_{21} U^d_{22}|^2=0.1269$, 
we find that the second term gives a severe constraint
$$
M_{22} > 99\ {\rm TeV} ,
\eqno(20)
$$ 
where we have set $|\Delta m_K^{fam}|_{max}=10^{-18}$ TeV
optimistically.

The most easy way to evade the constraint from $K^0$-$\bar{K}^0$  
is to assume $U^d \simeq {\bf 1}$. 
Then, the constraint from the $K^0$-$\bar{K}^0$ mixing 
disappears. 
However, then, 
we must consider $U^u = V_{CKM}^\dagger$ 
instead of $U^d = V_{CKM}$.
Then
, we will meet a similar problem on   
the observed $D^0$-$\bar{D}^0$ mixing: 
The $D^0$-$\bar{D}^0$ mixing gives
$$
\Delta m_D^{fam} = \left[\frac{1}{M_{33}^2} (U^{u *}_{31} U^u_{32})^2 +
\frac{1}{M_{22}^2}(U^{u *}_{21} U^u_{22})^2 +
\frac{1}{M_{11}^2}(U^{u *}_{11} U^u_{12})^2  \right]
\times {0.98974 \times 10^{-11} }\ {\rm [TeV]} ,
\eqno(21)
$$
where we have used (center values)
$f_D=0.2067$ GeV, $m_c(m_c)=1.275$ GeV and $m_u(2{\rm GeV})
 = 0.0023$ GeV. 
On the other hand, the present observed value \cite{PDG12}
is $\Delta m_D^{obs} = (8.38^{+2.8}_{-2.9}) \times 10^{-18}$
TeV.
If we again take the mixing $U^u$ given in Eq.(19), 
from values $|U^{u *}_{31} U^u_{32}|^2=2.979 \times 10^{-9}$,
$|U^{u *}_{21} U^u_{22}|^2=0.1269$ and 
$|U^{u *}_{21} U^u_{22}|^2=0.1269$, we again find that 
a constraint 
$$
M_{22} > 251\ {\rm TeV} ,
\eqno(22)
$$ 
where we have set $|\Delta m_D^{fam}|_{max}=2 \times 10^{-18}$ 
TeV. 

Of course, these constraints (20) and (22) are dependent of
the mixing values $U^d$ and $U^u$ and the setting of 
$|\Delta_{K,D}^{fam}|_{max}$,
so that those values  should be rigidly taken. 
Optimistically, we consider
$$
M_{22} \gtrsim 10^2 \ {\rm TeV}.
\eqno(23)
$$


\vspace{3mm}

\noindent{\large\bf 5. Search for another models}

As seen in the previous section, we have concluded 
that a mass of the gauge boson $A_2^2$ must be larger
than a few hundred TeV.
However, if we take, for example, $M_{22} = 300$ TeV, 
then Model I \cite{KY_PLB12} predicts 
masses $M_{23}=218$ TeV and $M_{33}=73$ TeV, which
are too large compared with the values (10) and (15),
respectively, so that we cannot see deviations from the 
$e$-$\mu$-$\tau$ universality in the tau and upsilon
decays.
If we adhere to the idea that the family gauge boson
effects are visible, we are obliged to abandon Model I.

The motivation in Model I is in the idea that the family 
gauge boson contribution cancels the logarithmic term
$\log m_{ei}^2$ in the QED radiative correction, so that
the characteristic of the model is that the gauge coupling constant
is related to the electroweak gauge coupling constants.
Therefore, we could discuss the gauge boson mass values 
explicitly in this paper.

Let us consider a minor change of Model I keeping
the idea of a model with an inverted mass hierarchy. 
In the model I, the mass relation (3) has been obtained
by the following mechanism: 
We assume a scalar $\Phi^\alpha_i$ which is 
$({\bf 3}, {\bf 3}^*)$ of U(3)$\times$U(3)$'$ families 
and whose VEVs 
$\langle \Phi \rangle$ give the charged lepton mass
matrix $M_e$ as 
$(M_e)_{ij} = k_e \langle \bar{\Phi}^i_\alpha \rangle
\langle {\Phi}_j^\alpha \rangle$ and  
$\langle {\Phi}_i^\alpha \rangle \propto \delta_{i\alpha}
\sqrt{m_{ei}}$. 
We also consider another scalar $\Psi^\alpha_i$ whose VEV  
dominantly gives family gauge boson masses (we have been
assume $|\langle \Psi \rangle| \gg |\langle \Phi \rangle|$)
and satisfies a relation 
$\langle \Psi \rangle \langle \Phi \rangle \propto {\bf 1}$.
Then we can obtain the gauge boson mass relation (3).
Similarly, if we introduce a scalar $Y^{ij}_e$ with 
$\langle Y_e^{ij} \rangle \propto \langle \Phi^i_\alpha \rangle 
\langle {\Phi}^j_\alpha \rangle$ 
[$\Phi^i_\alpha$ is $({\bf 3}, {\bf 3})$ of U(3)$\times$O(3)] 
and we assume a relation 
$\langle Y_e \rangle \langle\Psi\rangle \propto {\bf 1}$, 
we can obtain a gauge boson mass relation
$$
m^2(A_i^j) \equiv M_{ij}^2 = k \left( \frac{1}{m_{ei}} +
\frac{1}{m_{ej}} \right)^2 .
\eqno(24)
$$
which gives the mass ratios $M_{33}: M_{22}: M_{11}=
1: 16.82: 3.477 \times 10^2$.

Note that, in this revised model (Model II), 
the cancellation condition of the factor $\log m_{ei}^2$
is given by 
$$
\varepsilon_i +\varepsilon_0 = 
 \log\frac{m_{ei}^2}{m_{e3}^2} + \zeta_I 
\log  \left( \frac{M_{i1}^2 M_{i2}^2 M_{i3}^2 }{
M_{31}^2 M_{32}^2 M_{33}^2} \right)_I 
=  \log\frac{m_{ei}^2}{m_{e3}^2} + \zeta_{II} 
\log \left( \frac{M_{i1}^2 M_{i2}^2 M_{i3}^2 }{
M_{31}^2 M_{32}^2 M_{33}^2} \right)_{II} , 
\eqno(25)
$$
with $\varepsilon_i =0$.
Here, in the second term, only $m_{ei}$-dependent part 
is extracted. 
Since the gauge boson masses satisfy the relation
$$
\left( \frac{M_{ij}}{M_{33}} \right)^2_{II} 
= \left( \frac{M_{ij}}{M_{33}} \right)^4_I   ,
\eqno(26)
$$ 
the $\zeta$ parameter defined by Eq.(4) satisfies
$$
\zeta_{II} = \frac{1}{2} \zeta_I ,
\eqno(27)
$$

By these modifications for $g_F^2$ and $M_{ij}/M_{33}$, 
we obtain a revised value of $M_{33}$, 
$$
(M_{33})^{\tau}_{II} = 206^{+253}_{-52} \ {\rm GeV}, 
\ \ \ (M_{33})^{\Upsilon}_{II}= 79^{+92}_{-18} \ {\rm GeV},
\eqno(28)
$$ 
from the observed deviation (2) in the tau decays and 
from the observed deviation (12) in the $\Upsilon(1S)$ decays, 
respectively.
Such small values of $M_{33} \sim 10^2$ GeV cannot be ruled out 
from the current lower bound \cite{Tevatron05} by the 
$X\rightarrow \tau^+ \tau^-$ search at the Tevatron, 
because the production rate of $A_3^3$ is much 
smaller than that of the conventional $Z'$ boson. 
However, since Model II predicts, at most, $M_{22} \sim 20$ TeV 
even $M_{33} \sim 1$ TeV, the model cannot clear the constraint
(23) from the observed $K^0$-$\bar{K}^0$ and $D^0$-$\bar{D}^0$
mixings. 

If we adhere to the idea of the Sumino's cancellation mechanism
and the idea of an inverted mass hierarchy, we can consider 
a model with the following mass spectrum 
$$
M_{ij} = k_B \frac{1}{m_{ei} m_{ej}} .
\eqno(29)
$$
(Since this model is not a minor change of Model I, we 
call it as Model B hereafter.)
Note that as seen in Eq.(25), exactly speaking, Models I and II
cannot give $\varepsilon_i=0$ exactly (although they 
approximately satisfy the cancellation condition such as 
the charged lepton mass relation 
practically holds), while Model B with the mass spectrum (29) 
can exactly cancel the QED $\log m_{ei}^2$ term because of
$(M^2_{i1}M^2_{i2} M^2_{i3})/(M^2_{31}M^2_{32} M^2_{33})
=(m_{e3}/m_{ei})^6$.  
Since Model B can give the mass ratio $(M_{22}/M_{33})^2=8.00
\times 10^4$, we can clear the constraint (23) for $M_{33}
\sim 1$ TeV. 
For the predictions based on Model B, we list those in Table 1
together with the results in Models I and II. 
As seen in Table 1, Model B seems to be in favor of the observed 
values.
However, the big problem of Model B is that we cannot build
a model with such the family gauge boson mass spectrum (29) at 
present.

\begin{table}
\caption{
Numerical results in typical models.
Values of $M_{23}$ and $M_{33}$ have been
extracted from the observed deviations from
tau and upsilon decays, respectively.
Example values for $M_{23}$ and $M_{33}$
have been deduced from the lower value of $M_{22}$
which has been obtained from the observed $K^0$-$\bar{K}^0$ 
and $D^0$-$\bar{D}^0$ mixings (note that those values are
dependent on the value of $g_F$).  
The values of $M_{ij}$ are presented in a unit of TeV. 
}

\vspace{2mm}
\begin{tabular}{|c|ccc|} \hline
       & Model I & Model II & Model B \\ \hline
$\zeta$ & $1.752\equiv \zeta_I$ & $\frac{1}{2} \zeta_{I}$ &
$\frac{1}{3} = 0.1903 \zeta_I$ \\[.1in]
$\alpha_F$ & $0.022254$ & $ 0.01127$ & $0.004293$ \\[.1in]
$M_{33}:M_{23}:M_{22}$ & $1:2.98:4.10$ & $1:8.91:16.8$ &
$1:16.8:283$ \\[.1in]
$M_{23}^{\tau}$ & $2.6^{+3.2}_{-0.7}$ 
& $1.84^{+2.25}_{-0.46}$ & $1.13^{+2.25}_{-0.20}$ \\[.1in]
$M_{33}^\Upsilon$ & $0.112^{+0.130}_{-0.026}$ & $0.079^{+0.092}_{-0.018}$ 
& $0.049^{+0.056}_{-0.011}$ \\[.1in]
$M_{22}^{K,D}$ & $\gtrsim 300$ & $\gtrsim 200$ & $\gtrsim 130$ \\ \hline
Example & $M_{22}\equiv 300$ & $M_{22}\equiv 200$ & $M_{22}\equiv 130$ \\
        & $M_{23}= 218$ &  $M_{23}= 106$ & $M_{23}= 7.7$ \\
        & $M_{33}= 73$  & $M_{33}= 12$   & $M_{33}= 0.46$ \\
\hline
\end{tabular}
\end{table}


\vspace{3mm}

\noindent{\large\bf 6.  Concluding remarks}

In conclusion, it has been pointed out that the sign of the deviation 
from the $e$-$\mu$ universality in the tau decays suggests an 
existence of family gauge bosons with an inverted mass 
hierarchy. 
Stimulated this fact, we have investigated 
possible phenomenology of a specific model (Model I)
\cite{KY_PLB12} 
with an inverted mass hierarchy proposed by Yamashita and the author. 
Since the gauge coupling constant $g_F$ is not free parameter in this model,
the observed values of the deviations from the $e$-$\mu$-$\tau$ 
universality in the tau decays and upsilon decays can,
in principle, determine the family gauge boson masses:  
$M_{23} =2.6^{+3.2}_{-0.7}$ TeV and 
$M_{33}=0.112^{+0.120}_{-0.026}$ TeV, respectively.
Regrettably, at present, the data have large errors, so that we could not 
obtain a conclusive value of $M_{33}$. 

On the other hand, in the present model, the gauge bosons $A_i^j$ are in the 
mass-eigenstates on the family basis in which the charged 
lepton mass matrix is diagonal, while the quarks are, 
in general, not in the mass-eigenstates in the family basis, so that 
family-mixings $U^u \neq {\bf 1}$ and  $U^d \neq {\bf 1}$
appear.
We  have also investigated $K^0$-$\bar{K}^0$ and 
$D^0$-$\bar{D}^0$ mixings, because it is the biggest obstacle to
a family gauge boson model with a lower scale.
The observed values of $\Delta m_K$ and $\Delta m_D$ put 
a severe constraint for the mass $M_{22}$:  
$M_{22} \gtrsim 3 \times 10^2$ TeV.  
This constraint gives a conclusion that the deviations from 
the $e$-$\mu$-$\tau$ 
universality  are invisible in the tau decays and upsilon decays. 
(Since the upper values of the predicted gauge boson masses
become infinity if we take 1.3 $\sigma$ of the errors in the data,  
the conclusion $M_{22} \gtrsim 3 \times 10^2$ TeV does not 
contradict the results (10) and (15) considering their large errors, 
but such a case does not give
visible effects of the family gauge bosons.)

Since we want family gauge bosons whose effects are visible
in a lower scale physics, we have discussed alternative
model in Sec.5. 
Model II is a miner change of Model I, and it is possible to build 
such a model in fact, but the model cannot cope with both results,
that from the tau and upsilon decays, and that from  
 $K^0$-$\bar{K}^0$ and $D^0$-$\bar{D}^0$ mixings. 

Of course, there is an option that we abandon the Sumino's 
cancellation mechanism.
Then, the gauge coupling constant $g_F$ can become free parameter 
independently of the gauge boson mass spectrum, 
so that we can fit all data freely. 
However, in Sec.5, we have not taken such the option. 
We have inherited the ideas in the Sumino mechanism and in 
\cite{KY_PLB12}. 
If we abandon the Sumino mechanism, we will lose the motivation
to consider a family gauge boson model with the inverted mass 
hierarchy.
In Model B, the Sumino mechanism exactly holds (in Models I and II, 
the mechanism holds only approximately).
Model B can give interesting phenomenology, but we have not been able
to build such a model explicitly at present.
This is a future task to us.

We again would like to emphasis that if we improve 
the error values in the deviations from the $e$-$\mu$-$\tau$ 
universality in the tau decays and upsilon decays, 
we can determine the values of family gauge boson masses, 
i.e. the determination is within our reach.

If we leave the constraint from the  
$K^0$-$\bar{K}^0$ and $D^0$-$\bar{D}^0$ mixings, 
we can expect fruitful physics  not only in TeV but also 
sub-TeV regions.
For example, we expect a direct search for $A_3^3$  at the LHC.   
(For the details of the direct search for the lightest family gauge 
boson $A_3^3$ at the LHC, we shall report elsewhere.)
Very recently, an interesting decay model via a family changing 
neutral gauge boson has been pointed out \cite{Buras}. 
Although we have discussed Model I, II and B in Sec.5, those
are only examples. The essential idea is that the family gauge 
bosons have an inverted mass hierarchy. 
If we adopt a view of such the family gauge boson model 
with the inverted mass hierarchy, it will offer to us 
fruitful new physics experimentally and theoretically. 
Further studies are our future tasks.

 \vspace{3mm}

{\Large\bf Acknowledgments} 

A part of the present work was done at NTU, Taiwan. 
The author would like to thank members of the particle physics
 group at NTU and NTSC for the helpful discussions and their 
cordial hospitality, and,  
especially, X.G.~He for his interest in this work 
and inviting the author to NTU.
The author also thank T.~Yamashita and K.~Tsumura for 
their valuable and helpful conversations. 
He also thank I.~I.~Bigi for his interest in this work
and valuable conversations, and Y.~Kuno and J.~Sato for their 
helpful comments on the $\mu$-$e$ conversion.  
This work is supported by JSPS (No.\ 21540266).

\end{document}